\documentclass[preprint,review,12pt]{elsarticle}
\usepackage{amssymb}
\usepackage[numbers]{natbib}
\usepackage{lineno}
\usepackage{amsmath,environ}
\usepackage[utf8]{inputenc}
\usepackage[T1]{fontenc}
\usepackage{tabularx,ragged2e,booktabs,caption}
\usepackage{floatrow}
\usepackage{adjustbox}
\usepackage[titletoc,toc]{appendix}
\usepackage[flushleft]{threeparttable}
\setcitestyle{round}
\biboptions{authoryear, round}
\usepackage{soul}
\usepackage[margins]{trackchanges}
\setcitestyle{square,aysep={}}
\usepackage[nolists]{endfloat}
\usepackage[font=small,labelfont=bf]{caption}
\begin{document}
\begin{frontmatter}

\title{TSUFLIND-EnKF inversion model applied to tsunami deposits for estimation of transient flow depth and speed with quantified uncertainties}
\author[geovt]{Hui Tang\corref{cor1}}
\ead{tanghui@vt.edu}
\author[aoevt]{Jianxun Wang}
\author[geovt]{Robert Weiss}
\author[aoevt]{Heng Xiao}
\cortext[cor1]{Corresponding author}
\address[geovt]{Department of Geoscience, Virginia Polytechnic State University, United States}
\address[aoevt]{Department of Aerospace and Ocean Engineering, Virginia Polytechnic State University, United States}

\begin{abstract}

Tsunami deposits are recordings of tsunami events that contain information about flow conditions. Deciphering quantitative information from tsunami deposits is especially important for analyzing paleo-tsunami events in which deposits comprise the only leftover physical evidence. The physical meaning of the deciphered quantities depends on the physical assumptions that are applied. The aim of our study is to estimate the characteristics of tsunamis and quantify the errors and uncertainties that inherent within them. To achieve this goal, we apply the TSUFLIND-EnKF inversion model to study the deposition of an idealized deposit created by a single tsunami wave and one real case from 2004 India ocean tsunami. TSUFLIND-EnKF model combines TSUFLIND for the deposition module with the Ensemble Kalman Filtering (EnKF) method. In our modeling, we assume that grain-size distribution and thickness from the idealized deposits at different depths can be used as an observational variable. Our tentative results indicate that sampling methods and sampling frequencies of tsunami deposits influence not only the magnitude of the inverted variables, but also their errors and uncertainties. An interesting result of our technique is that a larger number of samples from a given tsunami deposit does not automatically mean that the inversion results are more robust with smaller errors and decreased uncertainties.
\end{abstract}

\end{frontmatter}

\section{Introduction}
The 2004 Indian Ocean tsunami killed over two hundred thousand people, left more than one million homeless and is the most destructive tsunami in human history so far \citep{JGRF:JGRF786}. However, as the population in coastal area increase quickly, future tsunamis can cause similar disasters or worse in the future \citep{Jaffe201290}. Since the first documentation of tsunami sediment research in the 1950s, many examples of modern and paleotsunami deposits have been reported \citep{Shepard1950,Goff2011107,Goff201270,Sugawara2014295}. Tsunami deposits play an important role not only in interpreting tsunami hydraulics, but also in tsunami hazard assessments \citep{Jaffe2007347,EOST:EOST16250,Goto201146,huitang2015}, as tsunami deposits are the only leftover physical evidence.

There are two major ways to study tsunami deposit by using numerical models: forward and inversion models. Unfortunately, due to the lack of pretsunami topography data, it is difficult to use forward models to study paleo tsunami. Inversion models become a very powerful and useful tool to identify the characteristics of paleo tsunami. The most prominent models for tsunami flow inversion are: Moore’s advection model \citep{Moore2007336}, Smith's model \citep{Smith2007362}, Soulsby’s model \citep{sousby2007}, TsuSedMod model \citep{Jaffe2007347} and TSUFLIND model \citep{huitang2015}. Many difficulties still remain in achieving accurate inversion results because of the intrinsic errors and uncertainties within tsunami deposit data set from tsunami field survey. A new method which can quantity the uncertainties will hold great promise for deciphering useful information from tsunami deposits \citep[see][]{Sugawara2014295,huitang2015}. Therefore, \citet{jianxun2015} propose a forward-inversion model framework (TSUFLIND-EnKF model) to estimate the flow condition and quantify the uncertainties in these estimations. The sedimentation module from TSUFLIND is employed for the deposition and the EnkF method is used for inverting hydrodynamics in TSUFLIND-EnKF model. In this study, we apply TSUFLIND-EnKF model and study the uncertainties in the final inversion results.

\section{Theoretical  Background}

A correctly defined inversion model consists of a forward model, inversion method and observational data. TSUFLIND-EnKF model's forward model is based on a simplified version of the TSUFLIND model. The EnKF method is used as an inversion technique. We employ the sediment flux as observational data to connect the forward model with the inversion method. 

\subsection{Forward Model}

In the TSUFLIND-EnKF model, TSUFLIND works as a forward model. TSUFLIND mainly incorporates three inversion models and adds some new modules to simulate tsunami deposit formation and estimate tsunami flow condition \citep{huitang2015}. The forward model starts with computing the grain-size distribution and sediment concentration. The grain-size distribution is assumed to be log-normal and characterized by the mean grain size, largest grain size and smallest grain size. The method employed to calculate the suspended sediment concentration is based on \citet{Madsen19931303}, and \citet{Jaffe2007347}. Another important assumption is that settling velocity is constant for given grain size during deposition process. The bed sediment concentration is computed with \citet{Madsen19931303}. The sediment flux at time step can be computed from average sediment concentration of grain size class at each water column layer. For more details about the forward model, we refer to \citet{jianxun2015} and appendix A. 

\subsection{EnKF Method}
The Ensemble Kalman Filtering (EnKF) method, firstly proposed by \citet{evensen1994}, is an extension of the Kalman Filtering method for solving nonlinear problems and widely used in weather prediction \citep{chui1999kalman}. As an inversion method, the EnKF method consists of three steps: initialization, propagation and analysis. Figure 1 depicts the flowchart outlining how this process works. 

\begin{figure}[!h]
  \centering
    \includegraphics[width=0.8\textwidth]{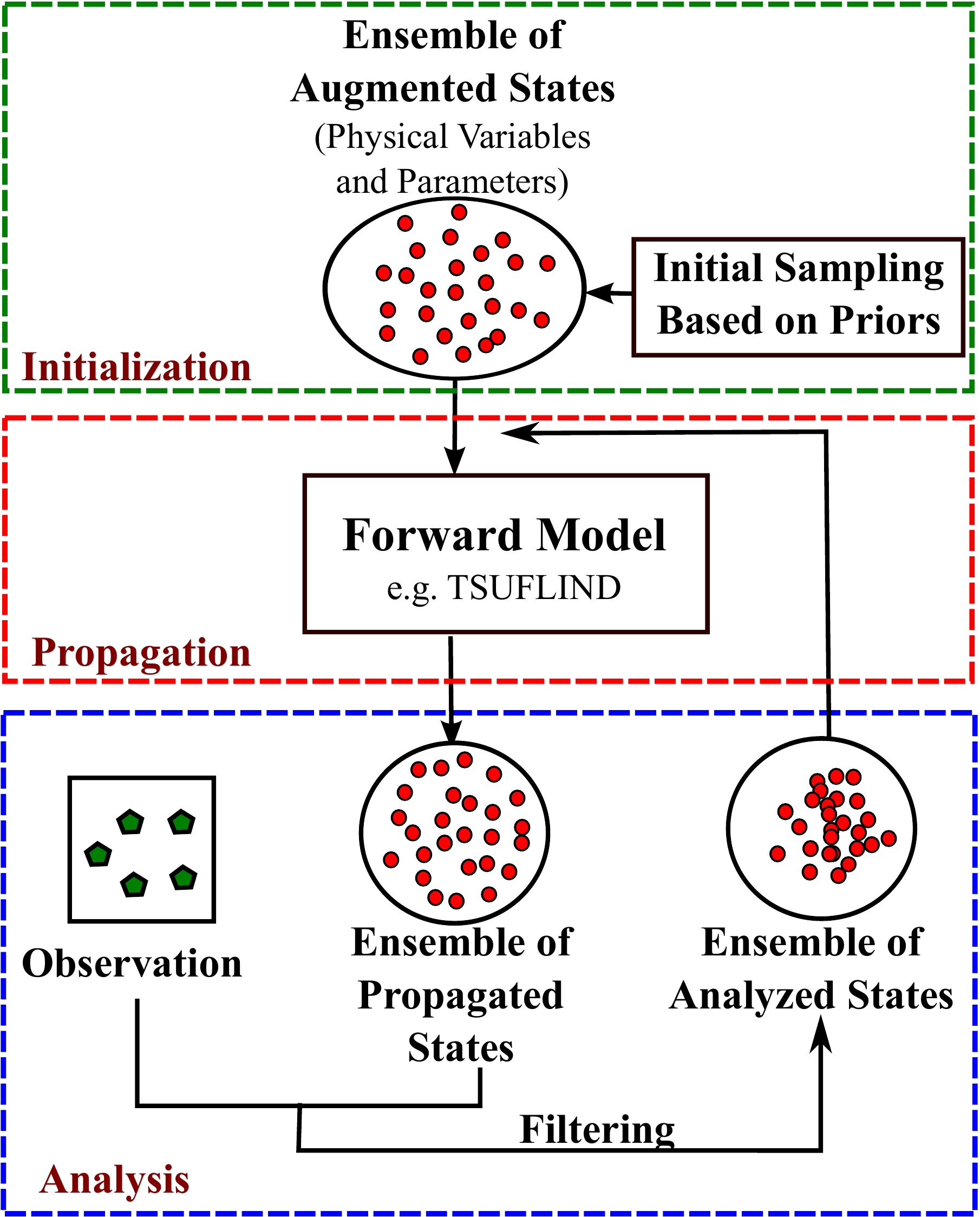}
  \caption{Flowchart for the EnKF method's iterative scheme to estimate tsunami flow parameters.}
\end{figure}

In the initial phase of the inversion process, we use the latin hypercube sampling method (LHS) to sample some initial ensemble for unknown variables based on prior. The LHS method is a series of statistical method for generating samples for parameters, which was described by \citet{mckey1979}. The unknown variables and observational variable are combined in an augmented state. Propagation is predicting the sediment flux based on the shear velocity and water depth with forward model. Propagation is performed for every $\Delta N$ forward model time steps which $\Delta N$ is the number of time steps between two data assimilation operations. After that, Kalman gain matrix can be calculated in the analysis part. Finally, the augmented state is renewed based on the observational data and the Kalman gain matrix. The propagation and analysis step will be repeated for next data assimilation operation to combine with observational data. The total number of operations depends on how much observational data is available. For more details about the EnKF method in TSUFLIND-EnKF model, we refer to \citet{jianxun2015} and appendix B.

\subsection{Inversion Result Evaluation and Error Model}
We utilize the relative error and second norm, $L_2$, to evaluate the inversion result and quantify inference error. For each ensemble, the inference error at time step $l$ is given by:

\begin{equation}
\varepsilon_{k}^l = \frac{{U}_{l,k}-\tilde{U}_{l,k}}{\tilde{U}_{l,k}}
\end{equation}
${U}_{l,k}$ and $\tilde{U}_{l,k}$ are the inverted value and synthetic truth for unknown variable $U$. The second norm at time step $l$ can be computed with:

\begin{equation}
{L_{2}^l}=\sqrt{\frac{\sum_{k=1}^{M}{\Big (\frac{U_{l,k}-\tilde{U}_{l,k}}{\tilde{U}_{l,k}}\Big)^2}}{M}}
\end{equation}
where $M$ is the number of ensemble. 

All experimental and modeled results, no matter how carefully measured and calculated, contain some errors. In order to study the influence of errors on the final inversion results, we propose a simplified error model that includes observational and model errors. In terms of observational error, there are two major sources: systematic error, $\beta$, and measurement error. Here we assume that the measurement error is equal to the measurement greatest possible error, $\epsilon$. The measurement's greatest possible error is usually half of the smallest scale of the experiment equipment. The observational error for the sediment flux is a relative error:

\begin{equation}
{E_{re}} = \frac{\beta \cdot \tilde{m}_{l}+\epsilon}{\tilde{m}_{l}}\cdot 100\%
\end{equation}
in which $\tilde{m}_{l}$ is the true mass of the tsunami sediment for sample $l$ in given location. Also, the entire sediment is uniformly divided to several sediment samples to simplify the framework, therefore we also can write $E_{re}$ as:

\begin{equation}
{E_{re}} = \beta +\frac{\epsilon}{m}\cdot{N_l}\cdot 100\%
\end{equation}
in which $m$ is the total mass of the tsunami sediment and $N_l$ represents the total number of sediment samples. It should be noted that error in the observation data increases with an increasing number of sediment samples. The model error is difficult to estimate, however, a constructed model error is applied.

\section{Results}
\subsection{Parameter Study}
The TSUFLIND-EnKF requires five inputs: number of ensemble, mean of ensemble, range of ensemble, type of unknown parameter (shear velocity and water depth) and of unknown parameters (uniform distribution or normal distribution). TSUFLIND-EnKF's sensitivity to these inputs is tested in a parameter study. Based on the parameter study, we hope to get some general conclusions about both TSUFLIND-EnKF model and inversion theory for tsunami research community.  The setting for all parameter studies can be found in Table 1. The effects of errors will be examined in detail in the section 3.2.

\begin{table}
\begin{threeparttable}
\caption{Physical and computational parameters used for the parameter study.}
\centering
\begin{adjustbox}{max width=\textwidth}
\begin{tabular}{l|c|c|c|c|c}
  \hline
  \textbf{Parameters}   & \textbf{Case 1}  &  \textbf{Case 2}  &   \textbf{Case 3} & \textbf{Case 4} & \textbf{Case 5} \\ 
  \hline
  Inferred parameters &   $u_*$        & $u_*$ & $u_*$  & $u_*$ & $u_*$ and $h$         \\ 
    \hline
  Grain size $\phi$ $(= - \log_2 [D/D_\textrm{ref}])$ &   \multicolumn{5}{c}{$0-3.25$}  \\
  $\Delta t$ & \multicolumn{5}{c}{0.5~s} \\
  Data assimilation interval $\Delta N$ & \multicolumn{5}{c}{10} \\
  std.~$\sigma_i$ of observation error  &
  \multicolumn{5}{c}{$\epsilon$ + 0.01$\tilde{\zeta}_i$} \\ 
   \cline{2-6}
  Number of samples $M$ & 10-3000 & \multicolumn{4}{c}{1000} \\
\cline{2-6}
Prior of unknown parameter& \multicolumn{3}{c|}{Uniform} & Uniform/Normal & Uniform \\
  \hline
  Synthetic truth $\tilde{u}_*$ for $u_*$ $ \mathrm{ ms}^{-1}$& \multicolumn{5}{c}{$0.5$} \\
   \cline{2-6}
    Mean of ensemble for $u_*$ ($\mathrm{ ms}^{-1}$) & 0.8 &0.25-1.0 &\multicolumn{3}{c}{0.8} \\
     \cline{2-6}
  Range of ensemble for $u_*$ ($\mathrm{ ms}^{-1}$) & \multicolumn{2}{c|}{0.2} & 0.1-1.6 & 0.8/- & 0.8 \\

  \hline
  Synthetic truth  $\tilde{h}$ for $h$ (m)&  \multicolumn{5}{c}{3}\\
  \cline{2-6}
  Range of ensemble for $h$ (m)& \multicolumn{4}{c|}{$-$} & 5\\
  Mean   $\bar{h}^0$ of ensemble for $h$ (m) &  \multicolumn{4}{c|}{$-$}  &   5 \\
  \hline
\end{tabular}
\end{adjustbox}
\end{threeparttable}
\end{table}
The sensitivity of the inversion result to the number of ensemble is tested by changing the number from 10 to 3000. The difference between various ensemble number's inference error $L_2$ norm decreases with more data assimilation operations taken (Fig. 2a). An increasing number of ensemble from 10 to 40 causes a decreasing the final inference error by 59\% (Fig. 2b). A decrease in the number of ensemble from 3000 to 40 results in the reduction of the final inference error by 63\% (Fig. 2b). 

\begin{figure}[!h]
  \centering
    \includegraphics[width=1\textwidth]{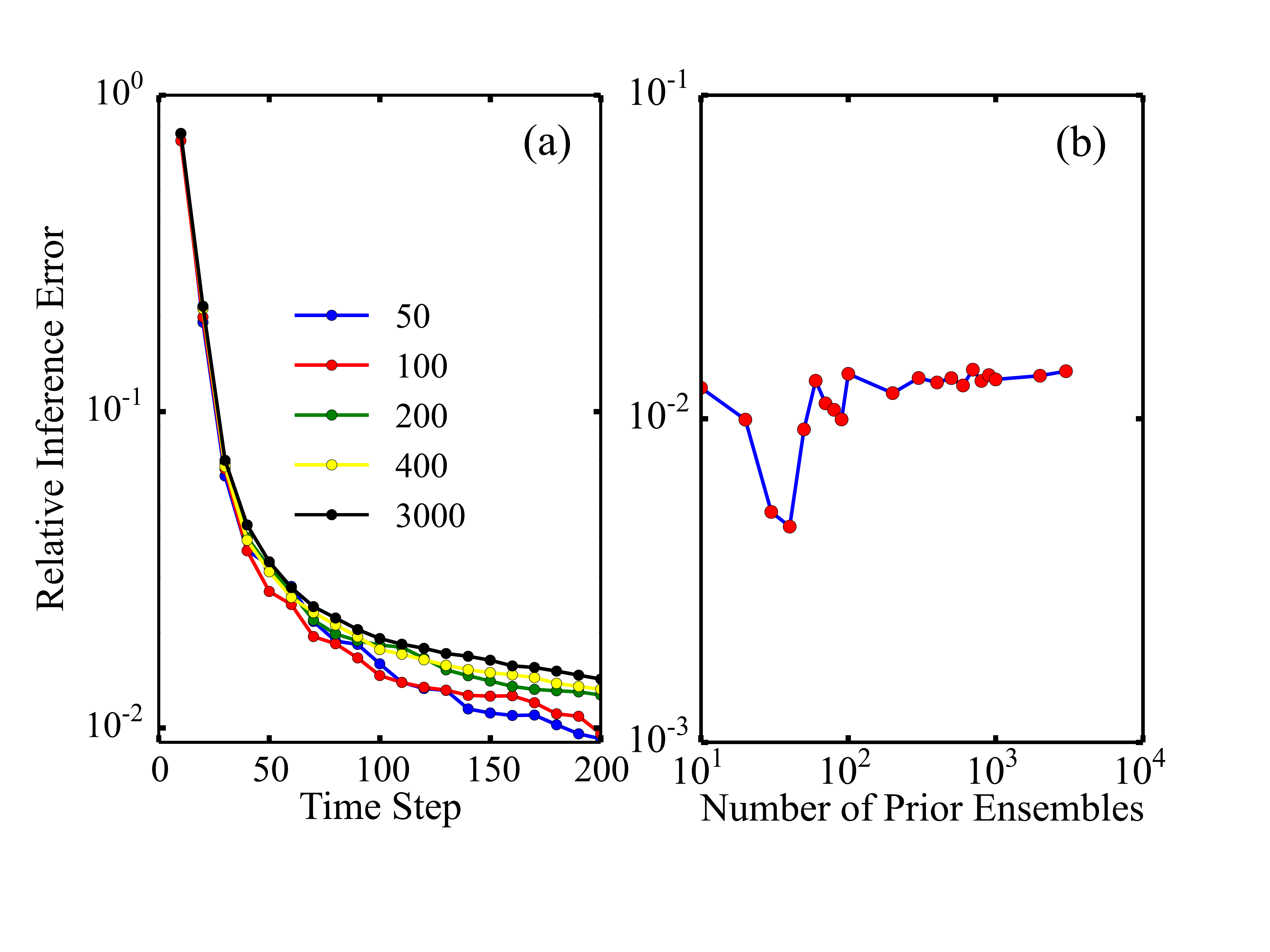}
  \caption{The relative inference error $L_2$ norm versus time steps for the number of ensemble from 10 to 3000. (a): The $L_2$ norm of inference error as a function of the time step and number of ensemble. The number of ensemble changes from 50 to 3000. (b): The final inference error $L_2$ norm as a function of the number of ensemble from 10 to 3000.}
\end{figure}

In order to study the influence of the mean value of ensemble, the mean value is adjusted from 0.25 to 1.0 $\text{ms}^{-1}$.  Figure 3a shows that the mean of ensemble will strongly change the inversion process pattern during inversion. However, the final inversion result is not a strong function of the mean value (Fig. 3b). For example, a 0.4 $\text{ms}^{-1}$ adjustment in the mean value from 0.4 $\text{ms}^{-1}$ to 0.8 $\text{ms}^{-1}$ results in only a 5\% change in the final inference error. The data resulting from a significant overestimation usually is superior to that of a dramatic underestimation. For instance, the inference error for the initial mean of 0.25 $\text{ms}^{-1}$ is 300\% larger than that from 0.75 $\text{ms}^{-1}$ (Fig. 3b). Also, the result distributions change from underestimation to slight overestimation when the mean of ensemble changes from 0.3 $\text{ms}^{-1}$ to 0.8 $\text{ms}^{-1}$ (Fig. 3c).

\begin{figure}[!h]
  \centering
    \includegraphics[width=1\textwidth]{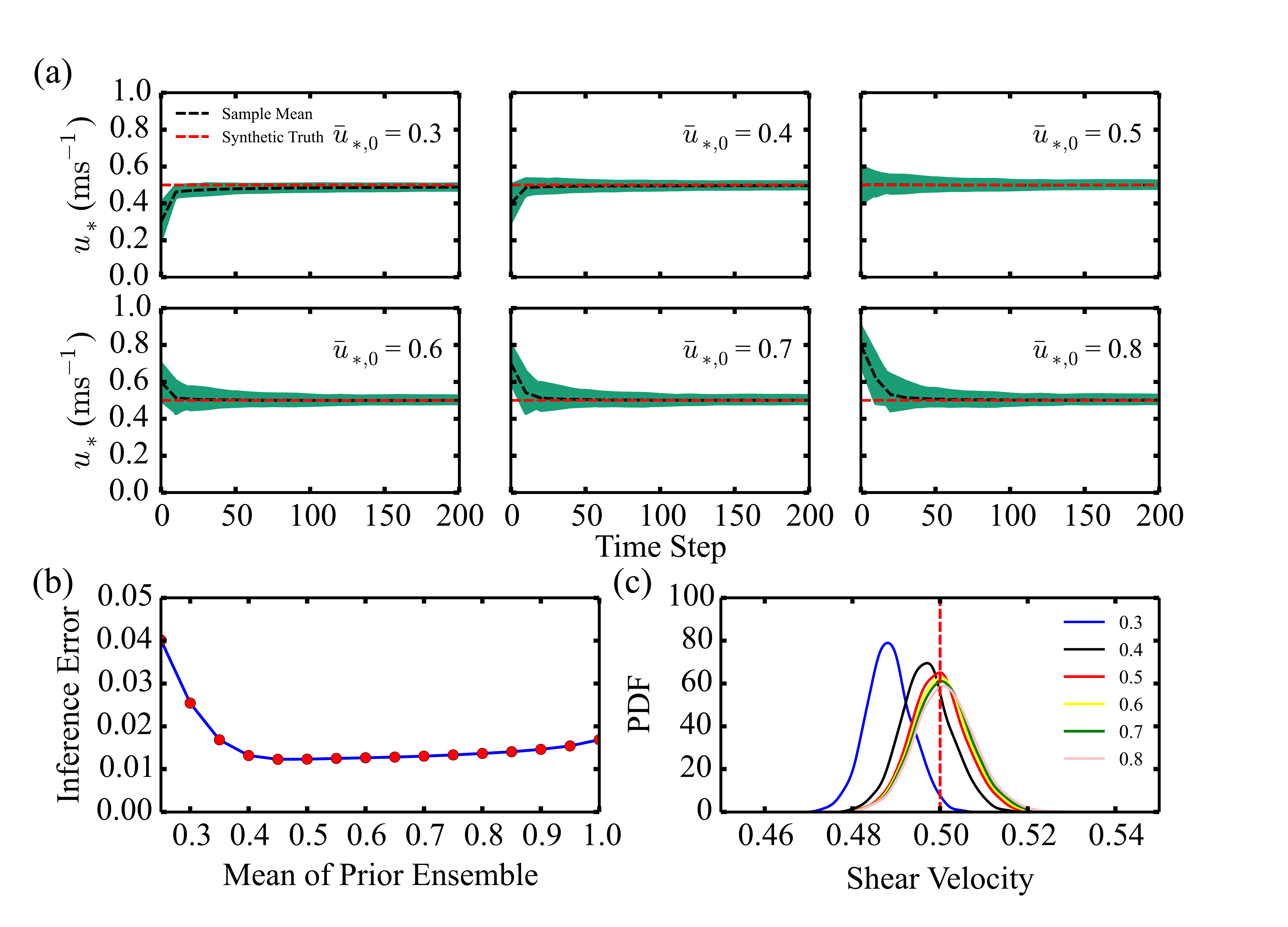}
  \caption{The shear velocity and $L_2$ norm of inference error as a function of the mean of ensemble ranging from 0.25 to 1.0 $\text{ms}^{-1}$. (a): The calculated shear velocity versus time step from 0 to 200 and mean of ensemble ranging from 0.3 to 0.8 $\text{ms}^{-1}$; The black dashed line is the averaged value of shear velocity for each time step. The Red dashed line is the synthetic truth for these cases. (b): The inference error $L_2$ norm of final inversion results versus mean of ensemble 0.25-1.0 $\text{ms}^{-1}$. (c): The final inversion result distributions for different mean of ensemble from 0.3 to 0.8 $\text{ms}^{-1}$. The Red dashed line is the synthetic truth for these cases.}
\end{figure}

The value range of ensemble will change the number of data assimilation operations needed to make results converge. Usually, the larger range requires more operations to convergent (Fig. 4a). The variance of the final inference error norm for range changing from 0.2 $\text{ms}^{-1}$ to 1.6 $\text{ms}^{-1}$ is only 9\% based on Fig. 4b. For the final result distribution, the final results follow a similar distribution pattern (Fig. 4c), which also indicates that TSUFLIND-EnKF is not sensitive to the range of ensemble unless the true value is not within the ensemble area.

\begin{figure}[!h]
  \centering
    \includegraphics[width=1\textwidth]{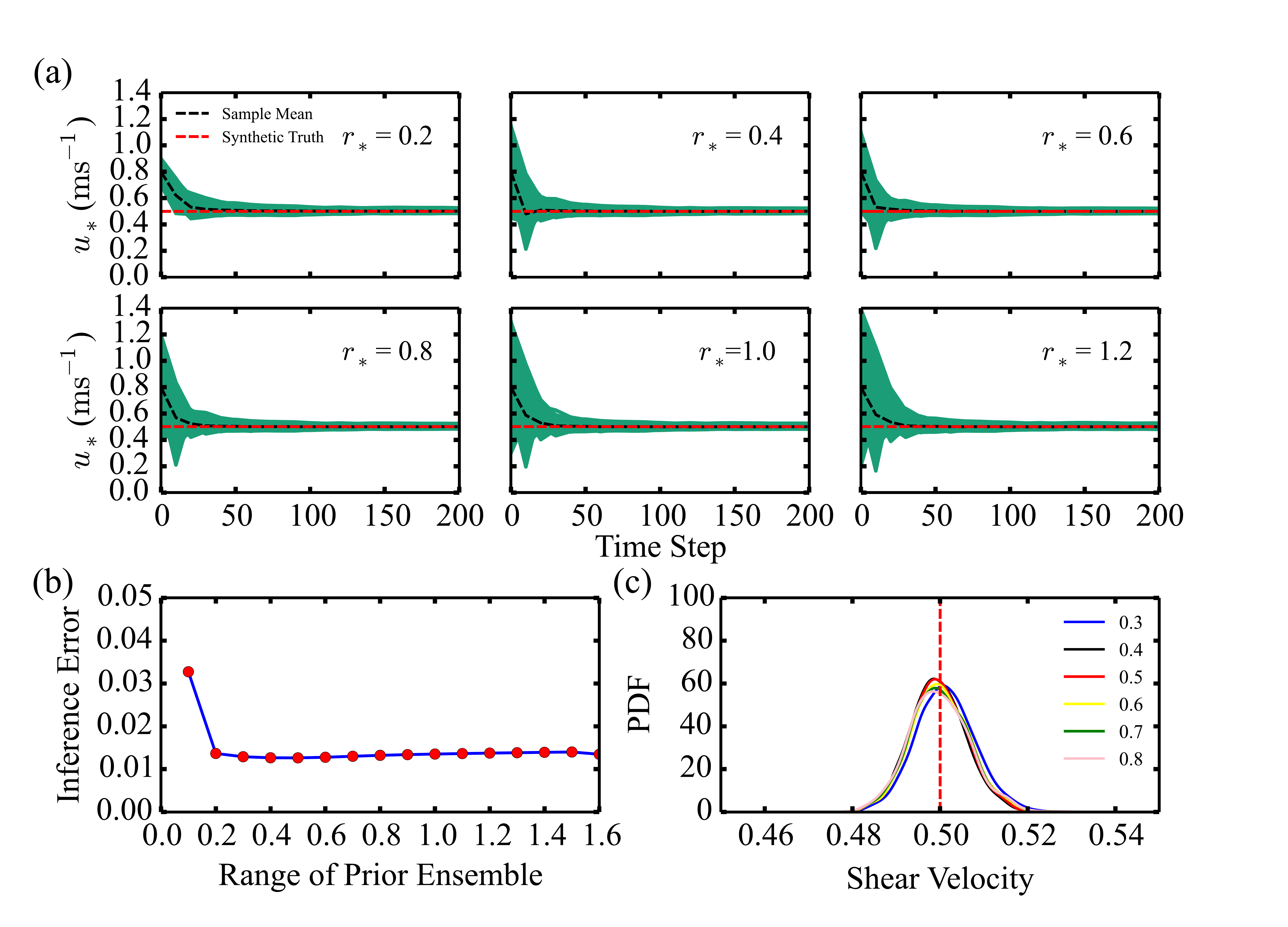}
  \caption{The shear velocity and inference error $L_2$ norm as a function of the range of ensemble ranging from 0.1 to 1.6 $\text{ms}^{-1}$. (a): The calculated shear velocity versus time step from 0 to 200 and range of ensemble ranging from 0.2 to 1.2 $\text{ms}^{-1}$; The black dashed line is the averaged value of shear velocity for each time step. The Red dashed line is the synthetic truth for these cases. (b): The inference error $L_2$ norm of final inversion results versus range of ensemble 0.1-1.6 $\text{ms}^{-1}$. (c): The final inversion result distributions for different range of ensemble from 0.3 to 0.8 $\text{ms}^{-1}$. The Red dashed line is the synthetic truth for these cases.}
\end{figure}

There are two major type of distribution for unknown parameter in TSUFLIND-EnKF: uniform distribution and norm distribution. We set up the model as shear velocity with 0.8 $\text{ms}^{-1}$ mean value, 0.8 $\text{ms}^{-1}$ range and 1000 ensemble. The range of ensemble for the uniform distribution is exactly from 0.4 to 1.2 $\text{ms}^{-1}$ (Fig. 5a). However, the normal distribution can only guarantee 68.2\% samples in the range from 0.4 to 1.2 $\text{ms}^{-1}$ (Fig. 5b). The distribution of unknown parameter strongly influences the inversion result distribution (Fig. 5c and 5d). The distribution has smaller interval in uniform distribution than normal distribution. However, the distribution of unknown does not influence the $L_2$ norm of the final inversion error significantly. The final $L_2$ norm by the normal distribution is only 9\% larger than the uniform distribution. 

\begin{figure}[!h]
  \centering
    \includegraphics[width=1\textwidth]{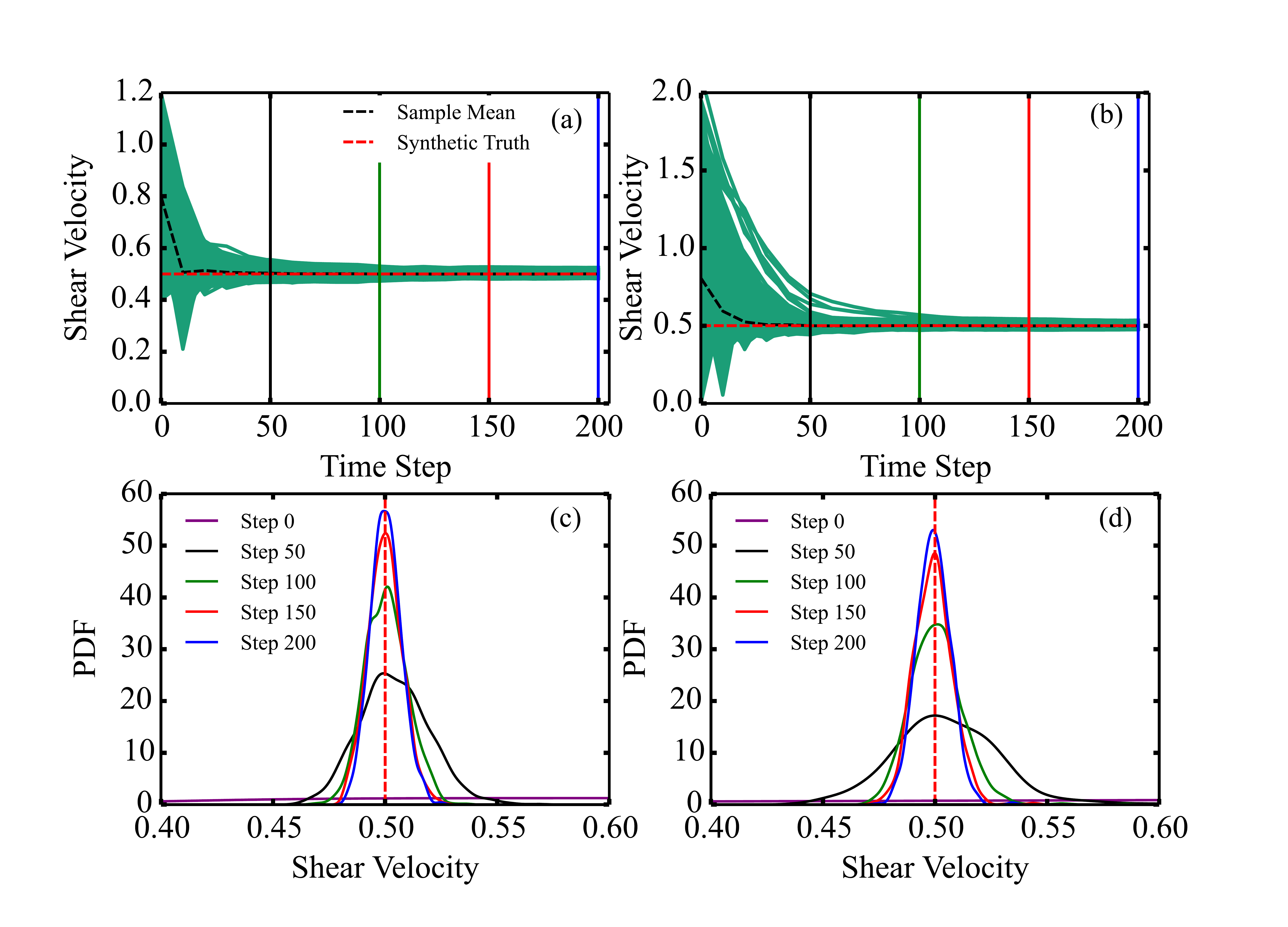}
  \caption{Compare two different of unknown parameter's inversion processes and results. (a): Tsunami flow shear velocity estimates for each time step by uniform distribution; (b): The shear velocity estimates for each time step by norm distribution; (c): the shear velocity distributions for time step 0, 50, 100, 150 and final result by uniform distribution; (d): the shear velocity distributions for time step 0, 50, 100, 150 and final result by norm distribution.}
\end{figure}

We only calculate shear velocity for previous parameter study. As a part of parameter study, we explore TSUFLIND-EnKF's sensitivity to the choice of unknown parameter. We choose flow depth and shear velocity as unknown parameters. It should be noted that the shear velocity and water depth are inverted separately in the first two cases (Fig. 6a, 6c, 6e and 6f). The flow depth and shear velocity inversion results for the same initial condition are shown in Fig. 6e and 6f. The norm of final inference error for water depth is 16.2\% larger than that for the shear velocity. In most real cases, both shear velocity and water depth are unknown. It should be noted that TSUFLIND-EnKF can estimate both water depth and shear velocity at the same time. In Figures 6b, 6d, 6g and 6h, the water depth and shear velocity are jointly inverted. The norm of final inference error for water depth is increased by 6\% when the combination inversion is applied. The norm of final inference error for shear velocity by separation inversion is 12\% smaller than that comes from combination inversion.

\begin{figure}[!h]
  \centering
    \includegraphics[width=1\textwidth]{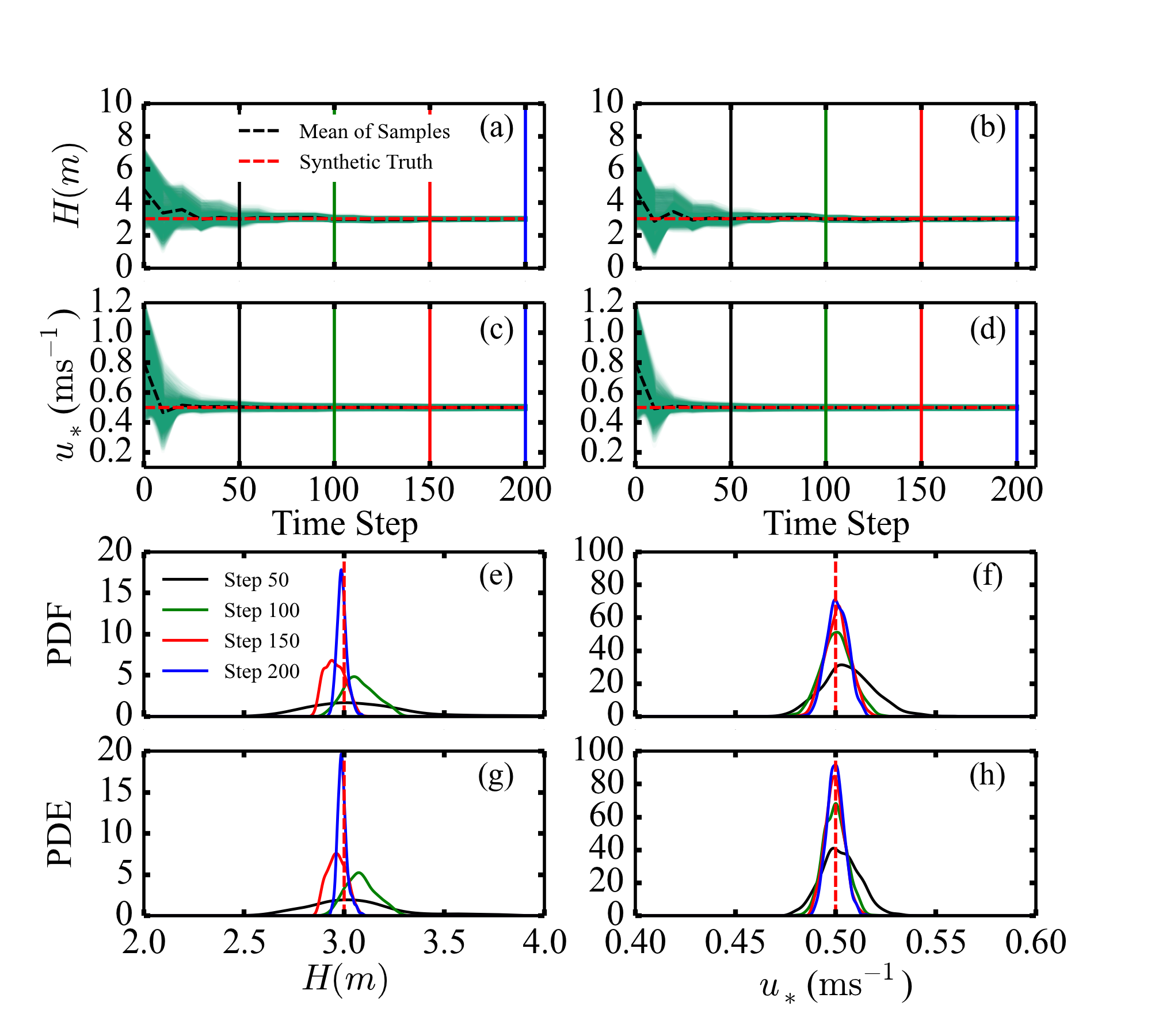}
  \caption{Compare two different unknown parameters inversion processes and results. (a)-(d): The inverted tsunami flow depths and shear velocities for each time step are indicated by the green area. The black dashed line is the average value of estimated unknown parameters. The red dashed line is the synthetic truth of estimated unknown parameters. (e)-(h): The water depth and shear velocity distributions for step 50, 100, 150 and final results. (a), (c), (e) and (f): Inverse water depth and shear velocity separately; (b), (d), (g) and (h): Inverse water depth and shear velocity together. }
\end{figure}

\subsection{Error Analysis}

The uncertainties of the inversion model arise from the fundamental theory employed and the observed data. We provide a systematic investigation of the uncertainty by varying the setting of observational data and the forward model. We anticipate a careful, but rigorous error analysis that can reveal important information and help to improve data collection during tsunami field survey.

This parameter study is carried by using an idealized tsunami deposit and provides the opportunity to study how observational and model errors influence the uncertainty of the final results. For this, it is assumed that all data from original idealized tsunami deposit and forward model are error free; the errors are added into the data manually. For Figures 7a and 7b, the model error is 1\% and the observational error ranges from 0.1\% to 30\%. The $L_2$ norm  of final inference error only increases by 6\%, when observational errors change from 0.1\% to 5\%. A change from 5\% to 30\% for the observational errors creates an obvious difference between these final reference error norms about 221.7\% (Fig. 7b). The observational error is 1\% and the model error changes from 0.1\% to 30\% in Fig. 7c and 7d. Based on Fig. 7a and 7c, it appears that the observational errors add more uncertainty to the inversion results than the model errors as long as both two types of errors are smaller than 10\%. When  both model errors and observational errors are larger than 10\%, both model errors and observational errors have a similar influence on the final inversion results. For the model error, the error norms of final inversion results decrease first and then increase again when the model error changes from 0.1\% to 30\% (Fig. 7d).

\begin{figure}[!h]
  \centering
    \includegraphics[width=1\textwidth]{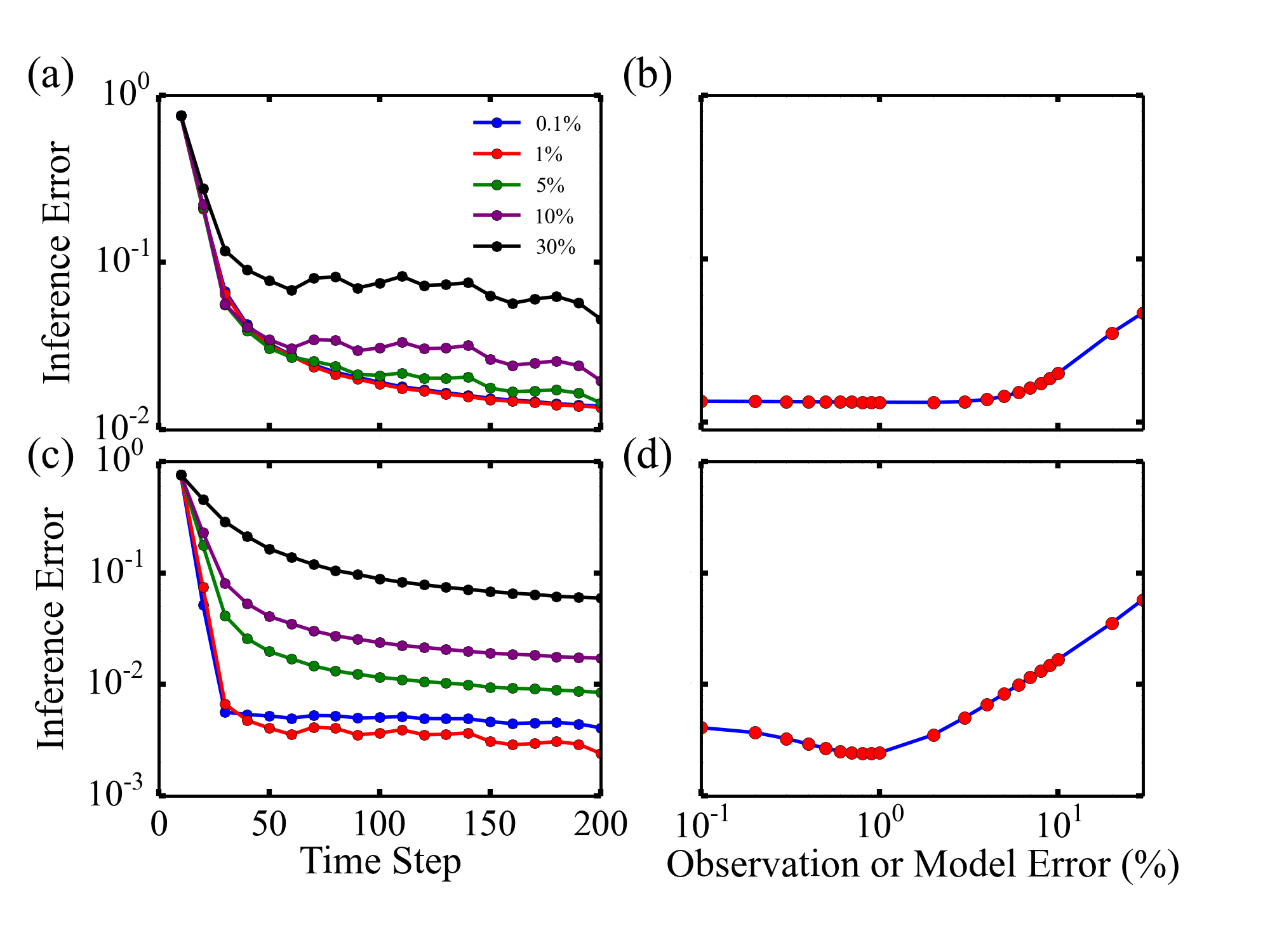}
  \caption{The $L_2$ norm of inference error as a function of time step and errors ranging from 0.1\% to 30\%. (a): The $L_2$ norm of inference error versus time step for observation error 0.1\% to 30\%. (b): The $L_2$ norm of inference error of final inversion results as a function of observation error from 0.1\% to 30\%. (c): The $L_2$ norm of inference error versus time step for model error 0.1\% to 30\%. (d): The $L_2$ norm of inference error of final inversion results as a function of model error from 0.1\% to 30\%.}
\end{figure}  

Estimating sediment flux during tsunami is complicated due to large fluctuation of hydrodynamic forces during tsunami flooding. For most cases, the sediment flux can be estimated with the help of tsunami deposits. In here, we assume that the sediment flux can be calculated by the total tsunami deposit thickness, sampling frequency (the number of sediment layers) and grain-size distribution. For the all samples, it is assumed that the observational error is constant in each sample. The error analysis for this case is based on field data and real cases. The total systematic error ($\beta$) and greatest possible errors ($\epsilon$) for thickness measurement are set as 5\% and 0.05 $\mathrm{mm}$ respectively. The inference error of final inversion result is sensitive to the number of sediment samples (Fig. 8a and 8b). For example, for a 30-cm-thick deposit, the final inference error $L_2$ norm for water depth is about 0.083 for nine sediment layers and 0.0003 for eleven sediment layers (Fig. 8a). The final $L_2$ norm of inference error decreases first and then increases. Based on Fig. 8a and 8b, the best frequency of sampling for shear velocity is 9 and for water depth is 11. 

\begin{figure}[h!]
  \centering
    \includegraphics[width=1\textwidth]{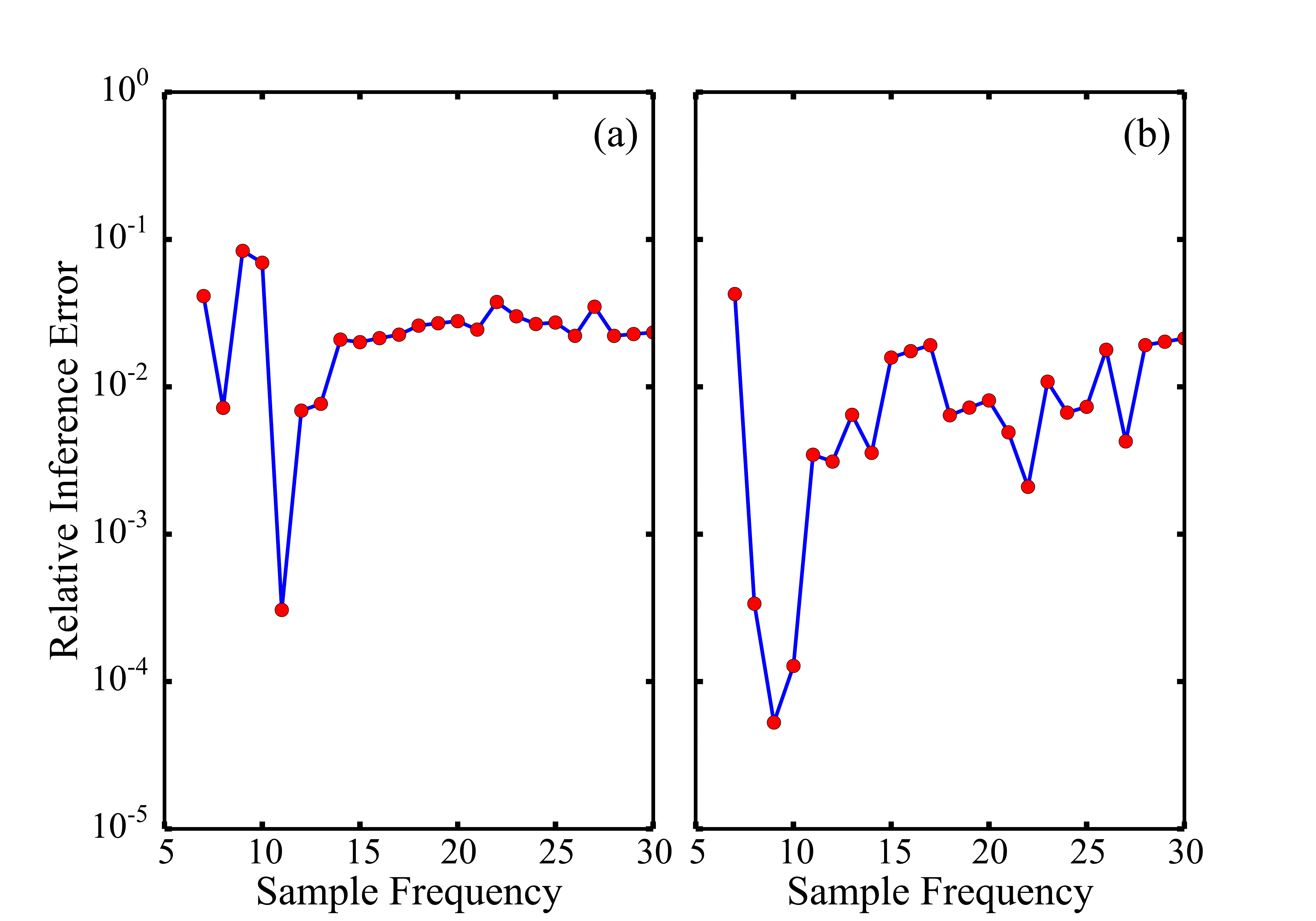}
  \caption{The final $L_2$ norm of inference error versus sample frequency and types of unknown parameters based on 30-cm-thick ideal tsunami deposit. (a): The final $L_2$ norm of inference error for water depth as a function of sample frequency from 6 to 30; (b): The final $L_2$ norm of inference error for shear velocity as a function of sample frequency from 6 to 30.}
\end{figure}

\subsection{Application and A Real Case}

In this section, we outline two applications based on real tsunami deposit. The data in this test case is from the coastal area in the vicinity of Ranganathapuram formed by 2004 India ocean tsunami \citep{Bahlburg2007}. They will be used to design the sample frequency during the field work (Fig 9a). We assume we can take sample every 2 cm, 2.5 cm, 3 cm, 3.5 cm, 4 cm, 4.5 cm to 5 cm in vertical direction, which controls the frequency of sampling for each sample location along the slope. When the sediment thickness is not thick enough for one sediment sample, these sediments are sampled as one. Based on the field data (thickness and grain size range) and sample frequency, we can carry out the similar error analysis shown in Fig. 8 to estimate the sample frequency with smallest inference error. Figure 9a shows the tsunami deposit thickness and the best frequency of sampling along the slope for inversion shear velocity. When tsunami deposit at 50 m from inland is taken five samples in vertical direction,  which suggests the sediment should be sampled every 2 cm, this location will get the lowest inference error. At the sediment extension, all tsunami sediments are sampled as one layer, so the inference error is a constant at this location. 

\begin{figure}[!h]
  \centering
    \includegraphics[width=1.0\textwidth]{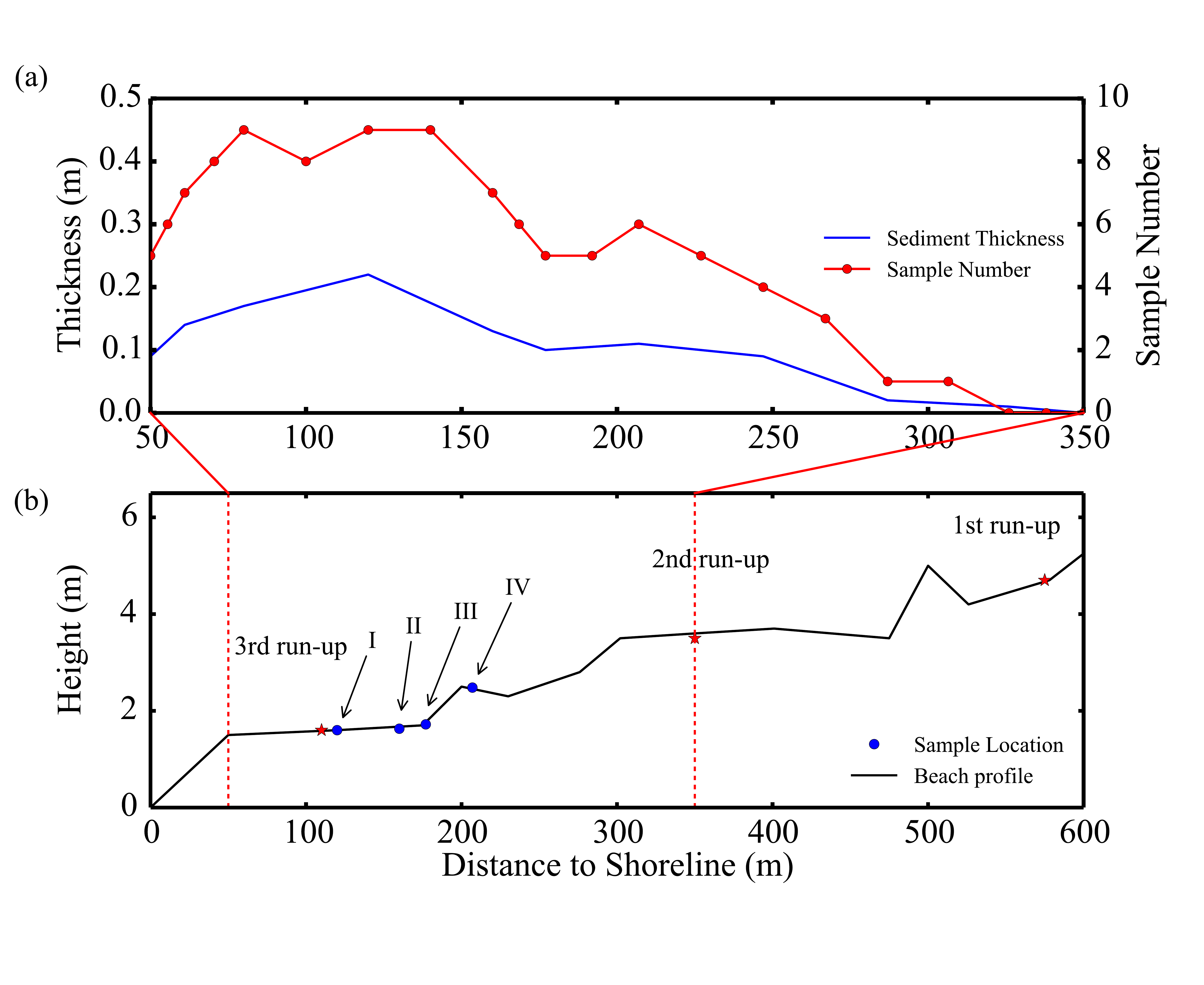}
  \caption{TSUFLIND-EnKF application cases for  2004 India ocean tsunami. (a): Sediment thickness and suggested sample number along the section in the vicinity of Ranganathapuram after 2004 India tsunami. The blue line is sediment thickness from field data. The red line with dot is the best sample number ; (b): The topography of Ranganathapuram cross section and sample location for test case. }
\end{figure}

In here, we apply TSUFLIND-EnKF to inverse shear velocity and flow depth and demonstrate the capability of our framework based on the best sample frequency. Then we calculate depth-averaged velocity and Froude number by using Eqs. A3 and A4. We choose four deposit samples from Ranganathapuram formed by 2004 Indian ocean tsunami (Fig 9b, $I$, $II$,$III$,$IV$). The sediment thickness for these four samples are 0.21 m, 0.12 m, 0.10 m and 0.09 m, respectively. The water depth and shear velocity are inverted together. For this test case, data assimilation operation is applied every ten time steps of the forward model. Figure 10 depicts the inversion results for the shear velocity, water depth, depth-averaged velocity and Froude number at four sampled locations along the slope. Unlike previous inversion models, TSUFLIND-EnKF presents the final inversion results as a probability density distribution function instead of only one value or range of values (Fig. 10). For the shear velocity, sample $I$ and $II$ have a similar range and distribution with a mean value of 0.159 $\text{ms}^{-1}$ (Fig 10 $I_a$ and $II_a$). And then shear velocity decrease to around 0.149 $\text{ms}^{-1}$ (Fig 10 $III_a$ and $IV_a$). The water depth decreases from the mean value of 2.905 to 2.502 m (Fig 10 $I_b$ to $IV_b$). The depth-averaged velocity has a similar pattern: decrease from 5.0 to 4.65 $\text{ms}^{-1}$. The mean Froude number does not change significantly in this case, which is around 0.94.

\begin{figure}[!h]
  \centering
    \includegraphics[width=1.1\textwidth]{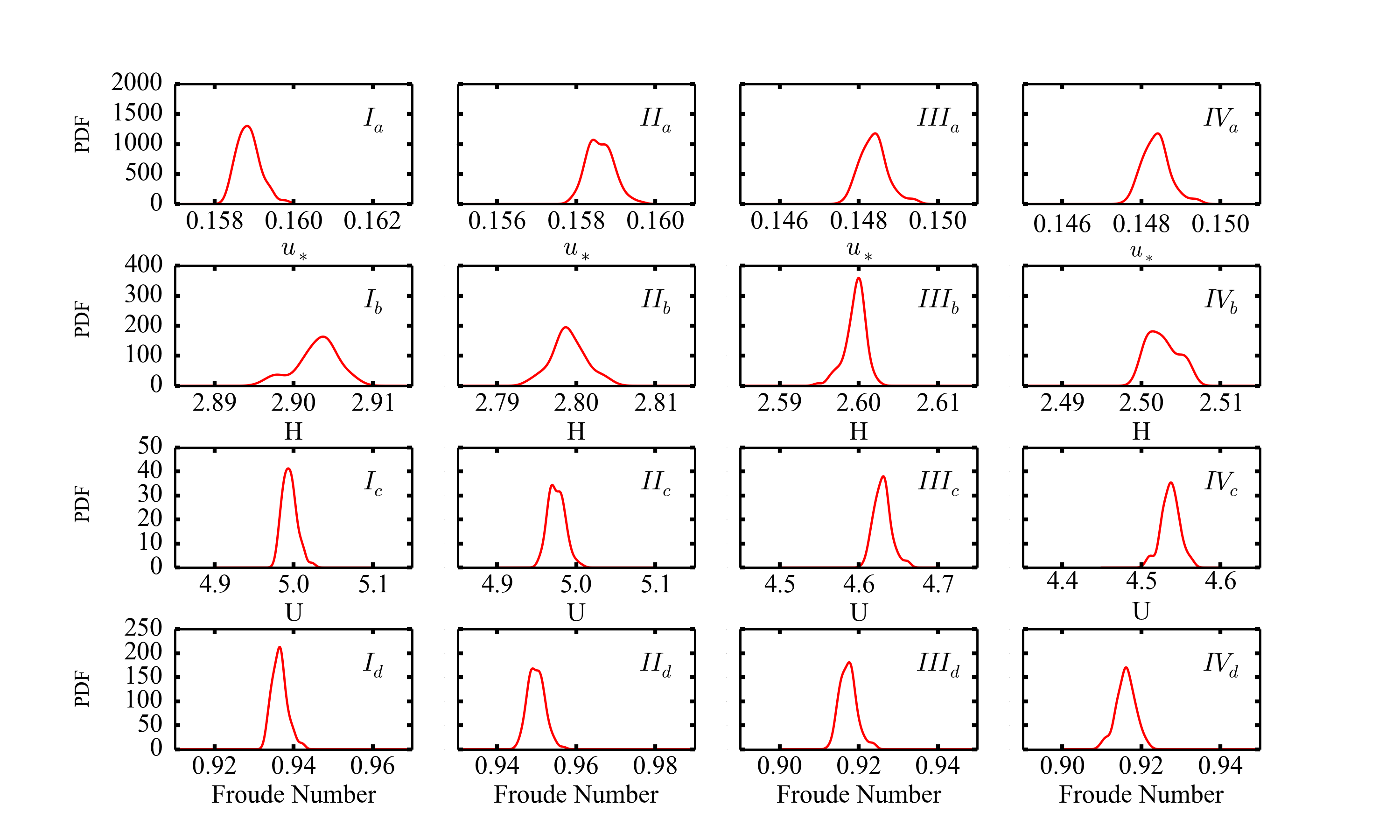}
  \caption{Inversion results for 2004 India ocean tsunami case from TSUFLIND-EnKF. $I_a$-$IV_a$: Estimates of shear velocity, $u_*$, for location $I$ to $IV$ in Fig 9b; $I_b$-$IV_b$: inversion results of water depth, $H$, for location $I$ to $IV$ in Fig 9b; $I_c$-$IV_c$: inverted depth-averaged velocity, $U$, for location $I$ to $IV$ in Fig 9b; $I_d$-$IV_d$: Froude number for location $I$ to $IV$ in Fig 9b.}
\end{figure}

\section{Discussion}

In this study, we apply TSUFLIND-EnKF model to estimate tsunami flow parameters and carry out parameter study. The TSUFLIND-EnKF model uses the TSUFLIND as a forward model and EnKF method for inversion. The EnKF method originally is developed for the situation that the mean of ensemble is the true value. A significant limitation is that the choice of ensemble (the mean, range and total number) may influence the final results in some extreme conditions. When TSUFLIND-EnKF model is applied, it is required that the mean of ensemble is close to the true value. Generally speaking, an overestimation of mean of ensemble does not significantly change the final inversion results, so the mean of ensemble can be set up a slightly larger than the results from other inversion models. Based on the parameter study, we find that there is an ideal number of initial guess ensemble to reduce the uncertainty in the inversion result (Fig. 2b). The distribution of unknown parameters usually will not significantly change the inversion result (Fig. 5). However, when the distribution of unknown parameters follows a given distribution, using given distribution in ensemble can help get good inversion results quickly. TSUFLIND-EnKF model is not applicable when the observation variable is not sensitive to unknown parameter. The parameter study suggests that the increment of unknown parameters in TSUFLIND-EnKF potentially decreases the quantity of the final inversion results (Fig. 6). In the same situation, different unknown parameters may need different number of data assimilation operations to converge. For instance, the shear velocity converges quicker than the water depth in the test case. The reason for this behavior is that the sediment flux, which is the observational variable, is more sensitive to the shear velocity than the water depth. 

For most cases, however, the sediment flux is almost impossible to measure during tsunami or other extreme high-energy events. Therefore, it has to be reconstructed by the deposit. The method to reconstruct tsunami sediment flux directly influences the uncertainty accommodate in the inversion results. When there are more errors in observational data from the deposit record than forward model, TSUFLIND-EnKF model will use more information from forward model results. Otherwise, TSUFLIND-EnKF will use more data from observational data. In other words, TSUFLIND-EnKF gets the best inversion results when the model error and the observational error are similar (Fig. 7). 

Also the application of high precision forward model would require higher precision observational data to get the inversion results with smaller uncertainty. For the sample frequency, the inference error decreases first and then increases with increasing sample frequency. This is because the more sediment samples provide more information to help calculate the unknown parameters. At the same time, increasing samples will add more errors into the model, so the uncertainty increases again (Fig. 8a and 8b). These results suggest that there might be a best field sample frequency, which significantly reduces the uncertainty in the inversion model results. Based on Fig. 8a and 8b, the best frequency of sampling for shear velocity is 9 and for water depth is 11. The insensitive unknown parameters like water depth usually need more sediment samples to get results with smaller inference error (Fig. 8a and 8b). 

When we apply specific forward model to EnKF method, the sampling frequency and measurement method become the major source of inference error. In order to reduce the inference error, the sample frequency and method should be decided before sampling during fieldwork. We apply the data from 2004 India ocean tsunami to design a sample frequency for this test case (Fig 9a). An interesting result of this application is that a larger number of samples from a given location does not automatically improve inversion results, because the more samples may input more observational errors into the system. For the second application, we estimated flow condition for four locations in Ranganathapuram during 2004 Indian ocean tsunami. Both depth-averaged velocity and Froude number are larger than results based on TSUFLIND. The mean flow velocity ranges from 5.0 to 4.65 $\text{ms}^{-1}$ with similar distribution, which is slightly larger than the result from TSUFLIND \citep{huitang2015} (4.7 to 4.5 $\text{ms}^{-1}$ from 150 m from shoreline to 207 m from shoreline) based on TSUFLIND model. One of the most important reason of the difference is that we directly apply the whole tsunami deposit, which may also consider bed load. Also, water depth used in \cite{huitang2015} is based on Moore's model and usually will overestimate as the linear assumption is applied. However, we can inverse water depth and shear velocity at the same time in TSUFIND-EnKF model that overcome the limitation of Moore's model. 

\section{Conclusion}
More field data is needed to quantify the limitation of the inversion model and how the tsunami flow can be better estimated. However, the TSUFLIND-EnKF model can provide meaningful information to improve estimates of these unknown parameters, such as speed and depth, and explore the inference error in the inversion results. Firstly, our study results show that TSUFLIND-EnKF gives reasonable estimates of unknown parameters (like shear velocity and water depth) for most conditions. Secondly, the sampling methods and frequencies of tsunami deposit influence not only the final inversion results, but also uncertainty the inversion result carry. The results of error analysis can work as a guide to help us improve future tsunami sediment field methods. From a more general viewpoint, these studies also indicate that the gap between results from modeling and fieldwork needs to be narrowed or closed because only together more robust inversion can be achieved. In this sense, researchers focusing on fieldwork must understand more about modeling, and more theoretical researchers must better comprehend the limitations and constraints of fieldwork. Also the application of a high-precision forward model would require high-precision data set.

\begin{appendices}
\section*{Appendix A: Forward Model}

The forward model based on \citet{huitang2015}, and \citet{Jaffe2007347} is briefly presented below. The sediment concentration is given by:
\begin{equation}
\label{eq:2}
C_{i}(z)=C_{i,0}e^{w_{i}\int_{z_0}^{z}{\frac{1}{K(z)}dz}}
\end{equation} in which $w_{i}$ is the settling velocity for grain size $i$, $z_0$ denotes the bed roughness and is calculated based on \citet{MacWilliams2005}. $z$ is elevation above sediment bed, and $k(z)$ represents the eddy viscosity profile from \citet{gelfenbaum1986}. 
In this equation, $C_{i, 0}$ denotes sediment concentration of the $i$\textsuperscript{th} grain-size class at the bed, which is calculated with \citet{Madsen19931303}:

\begin{equation}
\label{eq:7}
C_{i,0}=\frac{\beta_{0}(1-p)f_{i}S_{i}}{1+\beta_{0}S_{i}}
\end{equation}
where $\beta_{0}$ is the resuspension coefficient, $f_{i}$ is a fraction of the sediment of grain size $i$. $S_{i}$ is the normalized excess shear stress given by 

\begin{subequations}
\begin{align}
\label{eq:8}
S_{i}= \frac{\tau_{b}-\tau_{{cr}_{i}}}{\tau_{{cr}_{i}}} \:\:\: for \:\:\: \tau_{b} > \tau_{{cr}_{i}} 
\\
S_{i}= 0 \:\:\: for \:\:\: \tau_{b} \leq \tau_{{cr}_{i}}
\end{align}
\end{subequations}
where $\tau_{b}$ is the bed shear stress and $\tau_{{cr}_{i}}$ is the critical shear stress of the initial sediment movement \citep{Madsen19931303}.  The eddy viscosity profile, $k(z)$,  from \citet{gelfenbaum1986} is given by:

\begin{equation}
\label{eq:eddy}
K(z) = \kappa \, u_* \, z \exp\left[ {\frac{-z}{h} - 3.2 \left( \frac{z}{h} \right)^2 + \frac{2}{3} \times 3.2
    \left( \frac{z}{h} \right)^3} \right],
\end{equation}
where $\kappa =0.41$ is the von Karman constant, $z$ is the elevation above bed, $h$ denotes the water depth and $u_*$ is the shear velocity. After determining the eddy viscosity profile, the flow speed profile is calculated by :

\begin{equation}
\label{eq:6}
\ U(z)=\int_{z_0}^{z}{\frac{u_*^{2}}{K(z)}dz}
\end{equation}
where $z_o$ is the bottom roughness from \citet{MacWilliams2005}. The total time taken for all the sediment in the water column to settle, $T$, is divided to $N$ time steps of size $\Delta t$ such that $T = N \, \Delta t$. For grain size class $i$, the water column can also be divided to $N$ layers. The sediment in layer $l$ consists of the sediment in the $l$\textsuperscript{th} layer of the water column for all grain size classes, which thickness $\Delta \eta_l$ is computed by:

\begin{equation}
    \label{eq:d-eta}
    \Delta \eta_l = \frac{1}{C_0} \left( \sum_{i = 1}^{n} {\overline{C}_{i, l} \, \Delta z_{i}}
    \right), 
 \end{equation}
  in which $C_0$ is the total sediment concentration at the bed, $n$ is the number of grain size classes, and $\overline{C}_{i, l}$ is the average concentration of grainsize class $i$ in the water column layer $l$, which can be obtained by a simple integration:
  
  \begin{equation}
    \label{eq:cbar}
    \bar{C}_{i, l} = \frac{1}{\Delta z_i} \int_z^{z + \Delta z_i} \, C_i (z) \, dz.
  \end{equation}
 The sediment flux is given by:
\begin{equation}
\label{eq:flux}
\zeta_{i, l}  = \frac{\overline{C}_{i, l} \, \Delta z_{i}}{\Delta t}  \quad \textrm{or
   } \\
\zeta_{i, l}  =  \overline{C}_{i, l} \;  w_i  \quad \textrm{ in which }  \quad
  w_i = \frac{\Delta z_{i}}{\Delta t} .
\end{equation}
in which $\Delta z_{i}$ represents the water column layer thickness and $\Delta t$ denotes the forward model time interval. $\overline{C}_{i, l}$ can be calculated from sediment concentration profile $C_{i}(z)$.  Based on Eqs. A6, A10 and A11, the sediment flux at each time step can be calculated.\

\section*{Appendix B: Algorithm for Ensemble Kalman Filtering}

\label{app:enkf}

The algorithm of the ensemble Kalman filtering for inverse modeling is summarized below. For further more details about this algorithm, we refer to \citet{jianxun2015}. Given the prior distribution of the parameters to be inferred (shear velocity $u_*$ and flow depth $h$) and sediment flux observations with error covariance matrix $\mathbf{R}$, the following steps are performed:
\begin{enumerate}
\item \textbf{Initialization:} Generate initial ensemble $\{{\mathbf{x}_j}\}_{j = 1}^M$ of size $M$,
  where the augmented system state is:
  
  \begin{equation}
    \label{eq:ini-x}
    \mathbf{x} = [\zeta_1, \cdots, \zeta_n,  u_*, h]'
  \end{equation}

\item \textbf{Prediction:} 
  \begin{enumerate}
  \item Propagate the state from current state at time $t$ to the next assimilation step $t + \Delta
    T$ with the forward model TSUFLIND, indicated as $\mathcal{F}$,
 
 \begin{equation}
    \label{eq:forward}
    \hat{\mathbf{x}}_j^{(t+\Delta T)} = \mathcal{F} [ \mathbf{x}_j^{(t)} ]
  \end{equation}
  in which $\Delta T = \Delta N \Delta t$, indicating that the observation data is assimilated every
  $\Delta N$ time steps.

\item
  Estimate the
  mean $\bar{\mathbf{x}}$ and covariance $\mathbf{P}^{(n+1)}$ of the ensemble as:
  
  \begin{subequations}
    \begin{align}
      \bar{\mathbf{x}}^{(t+\Delta T)} = \frac{1}{M}\sum_{j=1}^N{\hat{\mathbf{x}}^{(t+\Delta T)}_j}    \label{eq:mean} \\
      \mathbf{P}^{(t+\Delta T)} = \frac{1}{M-1} \sum_{j = 1}^{N} {\left( \hat{\mathbf{x}}_j\hat{\mathbf{x}}_j' -  
          \bar{\mathbf{x}}\bar{\mathbf{x}}' \right)^{(t+\Delta T)}}
 \label{eq:cov} 
    \end{align}
  \end{subequations}
\end{enumerate}

\item  \textbf{Analysis:}
  \begin{enumerate}
  \item Compute the Kalman gain matrix as:
    
    \begin{equation}
      \label{eq:kalman-gain}
      \mathbf{K}^{(t+\Delta T)} = \mathbf{P}^{(t+\Delta T)} \mathbf{H}' (\mathbf{H} \mathbf{P}^{(t+\Delta T)} \mathbf{H}' + \mathbf{R})^{-1} 
    \end{equation}

  \item Update each sample in the predicted ensemble as follows:
        
        \begin{equation}
      \label{eq:update}
      \mathbf{x}_j^{(t+\Delta T)} = \hat{\mathbf{x}}_j^{(t+\Delta T)} + \mathbf{K}
      (\boldsymbol{\zeta}_j  - \mathbf{H} \hat{\mathbf{x}}_j^{(t+\Delta T)}) 
    \end{equation}
  \end{enumerate}

\item Repeat the prediction and analysis steps until  all observations are assimilated.
\end{enumerate}
\end{appendices}

\bibliographystyle{apa}
\bibliography{enkf}

\end{document}